\documentclass[11pt,twoside]{article}
\usepackage{asp2006}
\usepackage{lscape}
\usepackage{natbib}
\usepackage{graphicx}
\def\edcomment#1{\iffalse\marginpar{\raggedright\sl#1\/}\else\relax\fi}
\reversemarginpar

\begin{document}
\title{Molecular cloud formation and magnetic fields in spiral galaxies}
\author{C. L. Dobbs, D. J. Price,}
\affil{School of Physics, University of Exeter, Stocker Road, Exeter, UK}
\author{and I. A. Bonnell}
\affil{School of Physics \& Astronomy, North Haugh, University of St Andrews, St Andrews, Fife, UK}

\begin{abstract}
We present ongoing hydrodynamic and MHD simulations of molecular cloud formation in spiral galaxies. The hydrodynamic results show the formation of molecular gas clouds where spiral shocks compress atomic gas to high densities. The spiral shocks also produce structure in the spiral arms, provided the gas is cold ($<$ 1000 K). When both hot and cold components of the ISM are modeled, this structure is enhanced. Properties such as the clump mass spectra and spatial distribution will be compared from clouds identified in these simulations. In particular the multiphase simulations predict the presence of much more interarm molecular gas than when a single phase is assumed. We also discuss very recent results from galactic-scale MHD calculations. From observational comparisons of the magnetic and thermal pressure, magnetic fields are expected to be a major factor in explaining the dynamics of the ISM, from kpc scales to those of star formation. We describe the difference in structure of the spiral arms, and the evolution of the global magnetic field for a range of field strengths.
\end{abstract}

\section{Introduction}
Numerical calculations have now become a major component of research into 
molecular clouds, galactic dynamics and the ISM. Here we present a summary of recent results which examine the influence of spiral shocks on the ISM.

The presence of young stars in spiral arms is often associated with spiral shock triggering of star formation, following theoretical analysis of the gaseous response to spiral density waves \citep{Roberts1969,Shu1973}. Several observations show an increased star formation rate or efficiency in the spiral arms (e.g., \citealt{Cepa1990,Seigar2002}), indicative of spiral triggering. The primary objective of our numerical analysis however is to determine how the gas accumulates into giant molecular clouds (GMCs) in which star formation occurs e.g., through the agglomeration of smaller clouds or instabilities in the ISM. 
Furthermore, we discuss the degree of inter-am structure, the topic of several recent numerical calculations  \citep{Kim2002,Chak2003,Dobbs2006}. Such substructure, in the form of spurs and feathering, is commonly observed in spiral galaxies \citep{LaVigne2006}, and often associated with star formation.

\section{Calculations}
We perform global hydrodynamical and magneto-hydrodynamical (MHD) calculations of a 3D galactic disk. All the calculations are non-self gravitating and use the Smoothed Particle Hydrodynamics (SPH) code. The modifications adopted to incorporate magnetic fields are described in \citet{PB2007}. 
The gas is subject to a galactic potential which includes a 4 armed spiral component \citep{Cox2002}.
The calculations described here are all isothermal, although we investigate different gas temperatures and apply a two phase medium.

\section{Hydrodynamical Results}
We show the structure of a section of the disk from 3 purely hydrodynamic simulations in Figure 1. 
The panels show cases when the ISM is cold (50 K), warm ($10^4$ K) and a two phase medium consisting of half 100 K and half $10^4$ K gas. 
Where the gas is warm ($10^4$ K), there is no substructure in the disk, and the spiral arms are smooth and continuous. On the other hand where the gas is cold, the spiral arms are clumpy, with substructure along the trailing side of the arms. At later times in the simulation, these clumps produce more prominent, widely spaced spurs \citep{Dobbs2006}.
The formation of the clumps is due to the dynamics of the spiral shock. Gas from initially dissimilar radii agglomerates into clumps where orbit crowding occurs in the spiral arms \citep{DBP2006}. Furthermore the passage of the ISM through the shock induces a velocity dispersion in the gas, providing the gas is clumpy, which is the case when there is cold gas \citep{Dobbs2006}.

The molecular hydrogen column density is also overplotted on the total column density on Figure~1.
The density of H$_2$ was determined post-process using an algorithm from \citet{Bergin2004}. When the gas is $\sim$100 K, the amount of molecular gas in the disk is 10\%. However for the simulation with warm gas, no molecular hydrogen forms, and the gas remains atomic. This is because the density of gas in the shock is too low for H$_2$ to form, and any H$_2$ that does form is immediately photodissociated. For the case where the gas is 50 K, the molecular gas is located in highly structured clouds, predominantly along the spiral arms. For the two phase result, the molecular clouds tend to be smaller, since the extra pressure from the warm gas confines the cold gas to smaller clumps. There is more H$_2$ in the inter-arm regions, since the clumps are more dense, and less readily photodissociated. Consequently twice as much molecular gas is present compared to the same time frame of the single phase simulation of cold gas. The difference between the simulations is also reflected in the cloud mass spectrum. For the two phase results, there are more lower mass clouds and the mass spectrum is $dN/dM \propto M^{-2.35}$ compared to $dN/dM \propto M^{-1.7}$ for a single phase medium \citep{Dobbs2007}.
\begin{figure}
\centerline{
\includegraphics[scale=0.24]{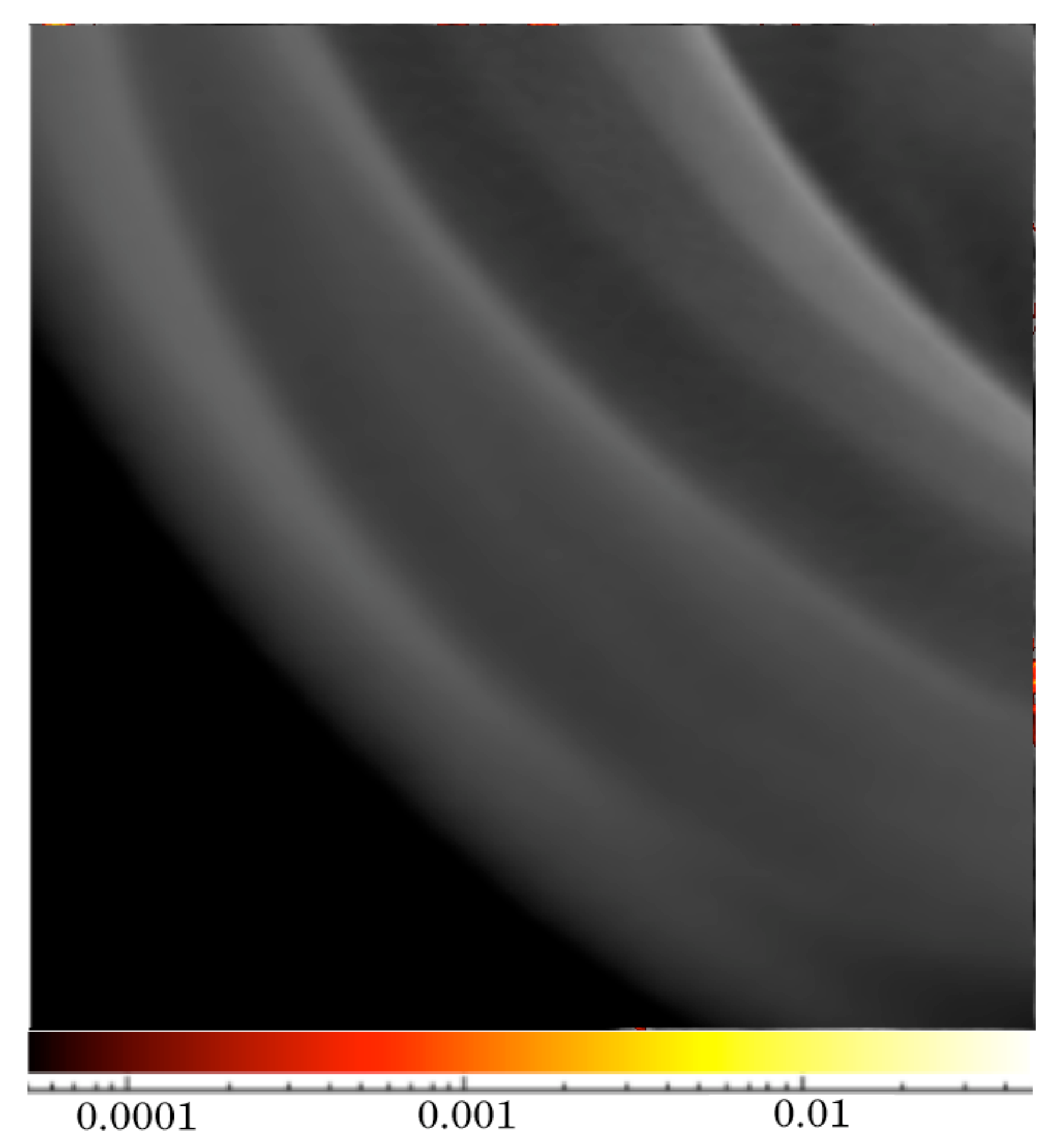}
\includegraphics[scale=0.24]{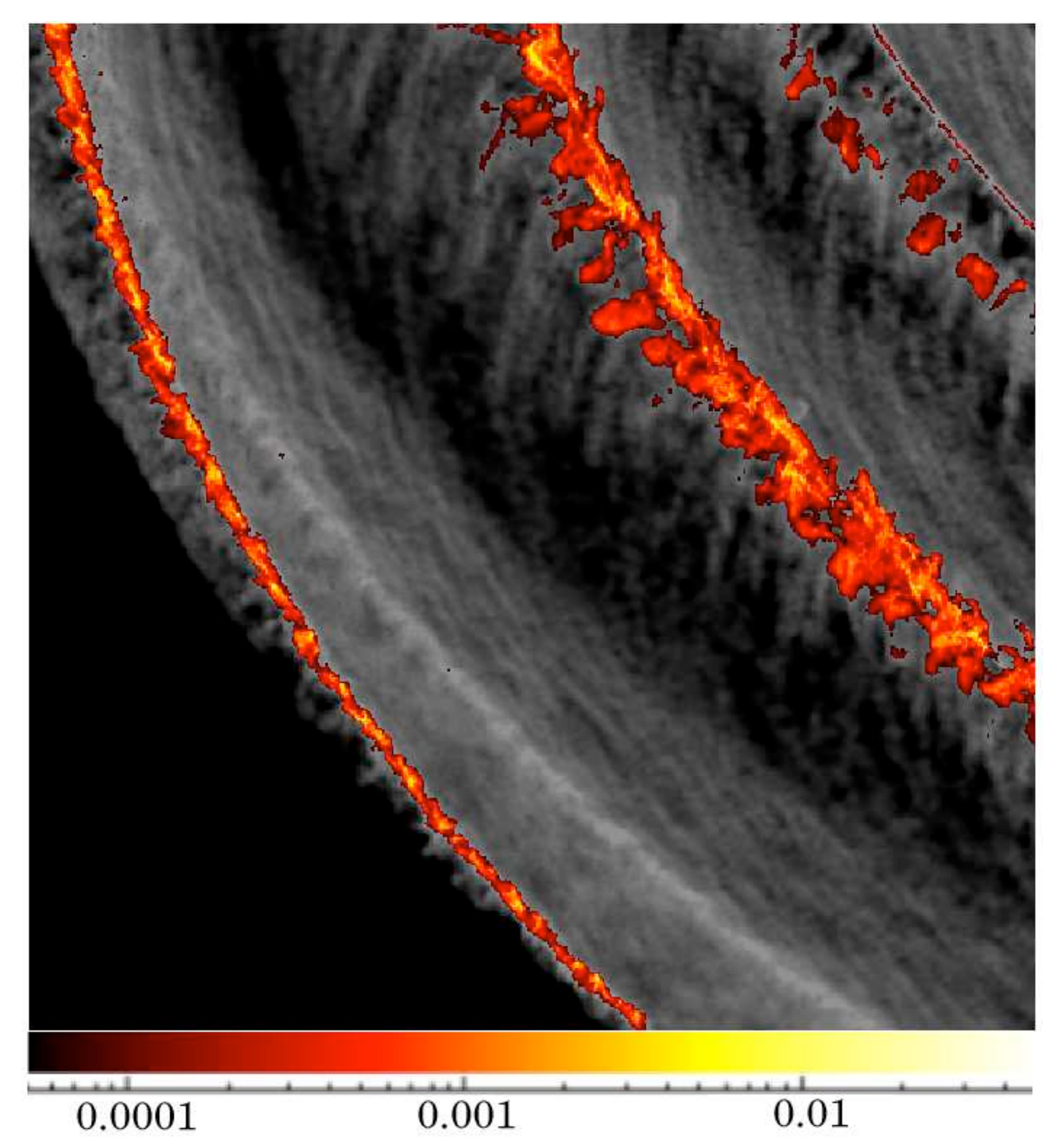}
\includegraphics[scale=0.24]{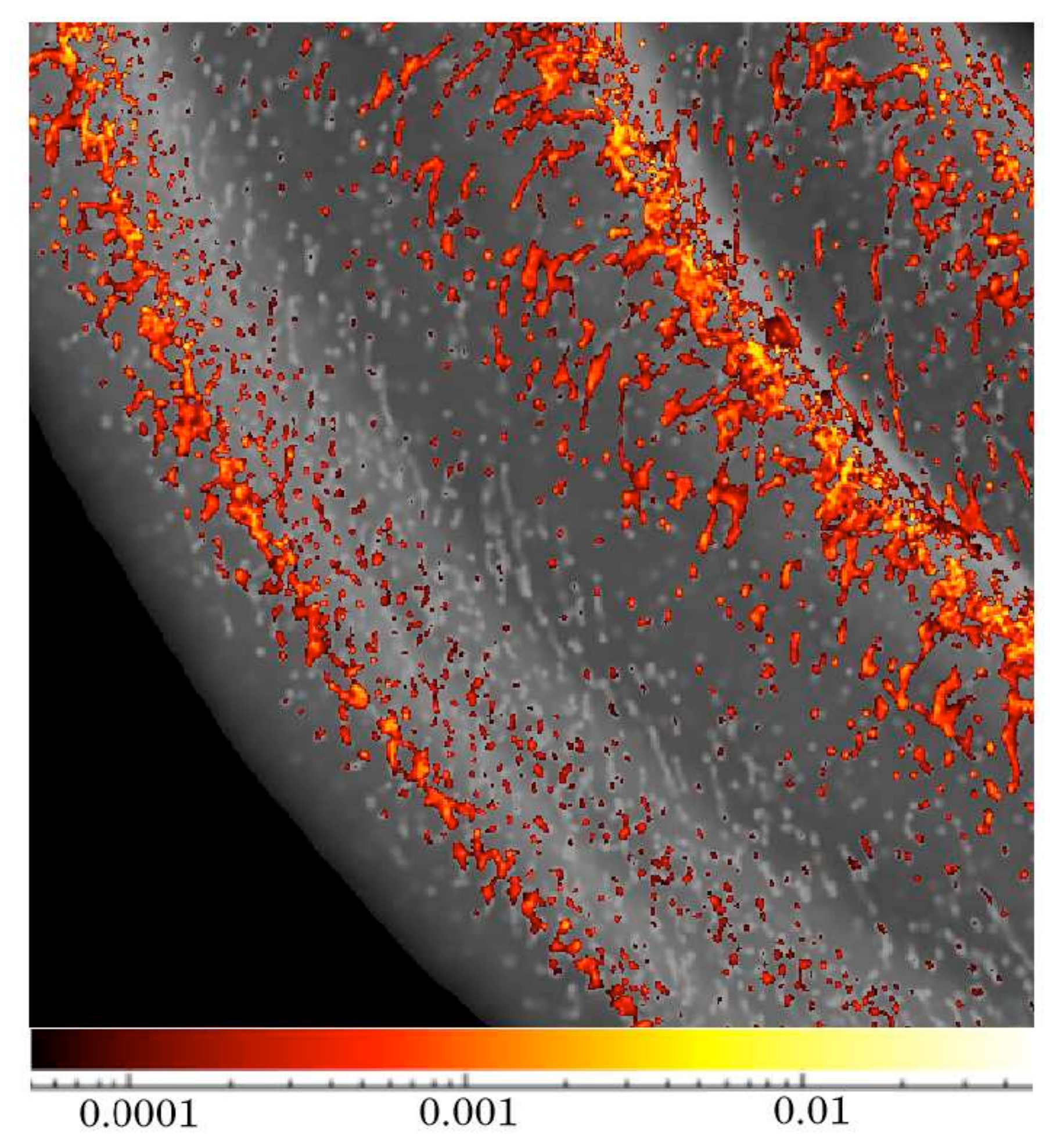}}
\caption{The spiral arm structure is shown for a 5 kpc x 5 kpc subsection of the galactic disk. The gas is 50 K (left), $10^4$ K (middle) and a two phase medium of half 100 K and half $10^4$ K (right). The scale indicates the column density of molecular gas (in g cm$^{-2}$), which is overplotted in red and yellow (see online version). For the $10^4$ K case, no molecular hydrogen is present. These panels primarily show that when cold gas is present, this component of the ISM is organised into clumpy structures by the spiral shock, which contain a substantial fraction of H$_2$.}
\end{figure}
\section{MHD Results}
In Figure~2, we show a section of the disk for different magnetic field strengths. For all the panels, the temperature of gas in the disk is 100 K. The shock becomes weaker as the magnetic field strength increases, and hence the structure in the disk decreases. The spiral arms also appear wider. Thus increasing the magnetic field has a similar effect to increasing the temperature in reducing the strength of the shock and smoothing out structure. Similar to the hydrodynamic results, we found that when the gas is warm ($10^4$ K), there is no substructure or inter-arm spurs. However when the gas is cold, substructure still occurs in the disk for a realistic value of $\beta$ (where $\beta$ is the ratio of the gas to magnetic pressure). 

The nature of the magnetic field also changes for different temperatures and field strengths. When the temperature is high, the field is very regular. However when the gas is cold, the magnetic field contains a substantial random component. This random component is associated with the velocity dispersion induced by the spiral shock.
Observations of the diffuse (warm) ISM indicate that the magnetic field contain both regular and disordered components, which are comparable in magnitude \citep{Beck2007}. In two phase simulations we find that the magnetic field of the warm gas has similar random and regular components \citep{DP2007}. The magnetic field in the cold gas is more disordered, although currently there are no corresponding observations of the degree of order for cold HI.   
\begin{figure}
\centerline{
\includegraphics[scale=0.24,angle=270]{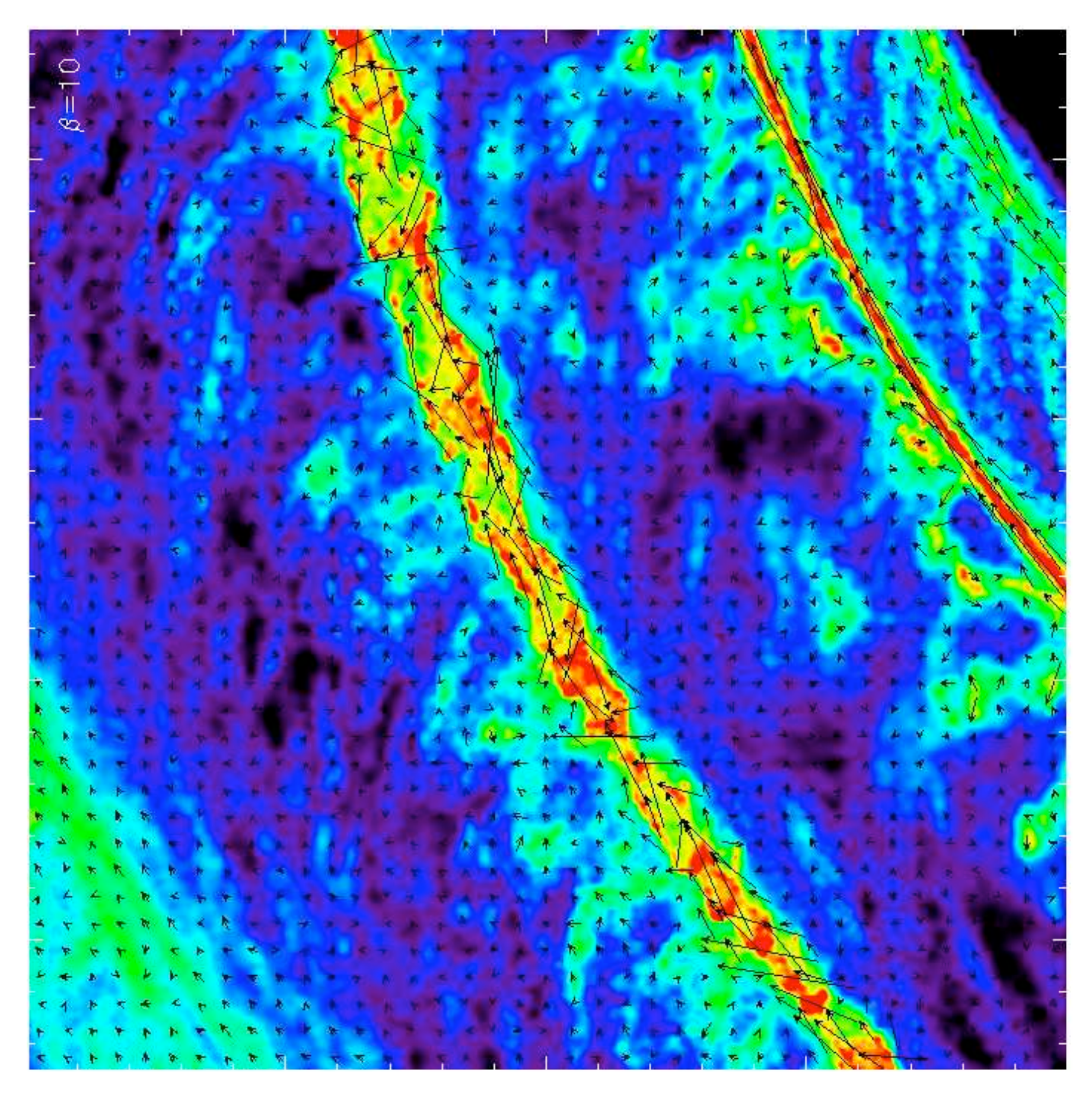}
\includegraphics[scale=0.24,angle=270]{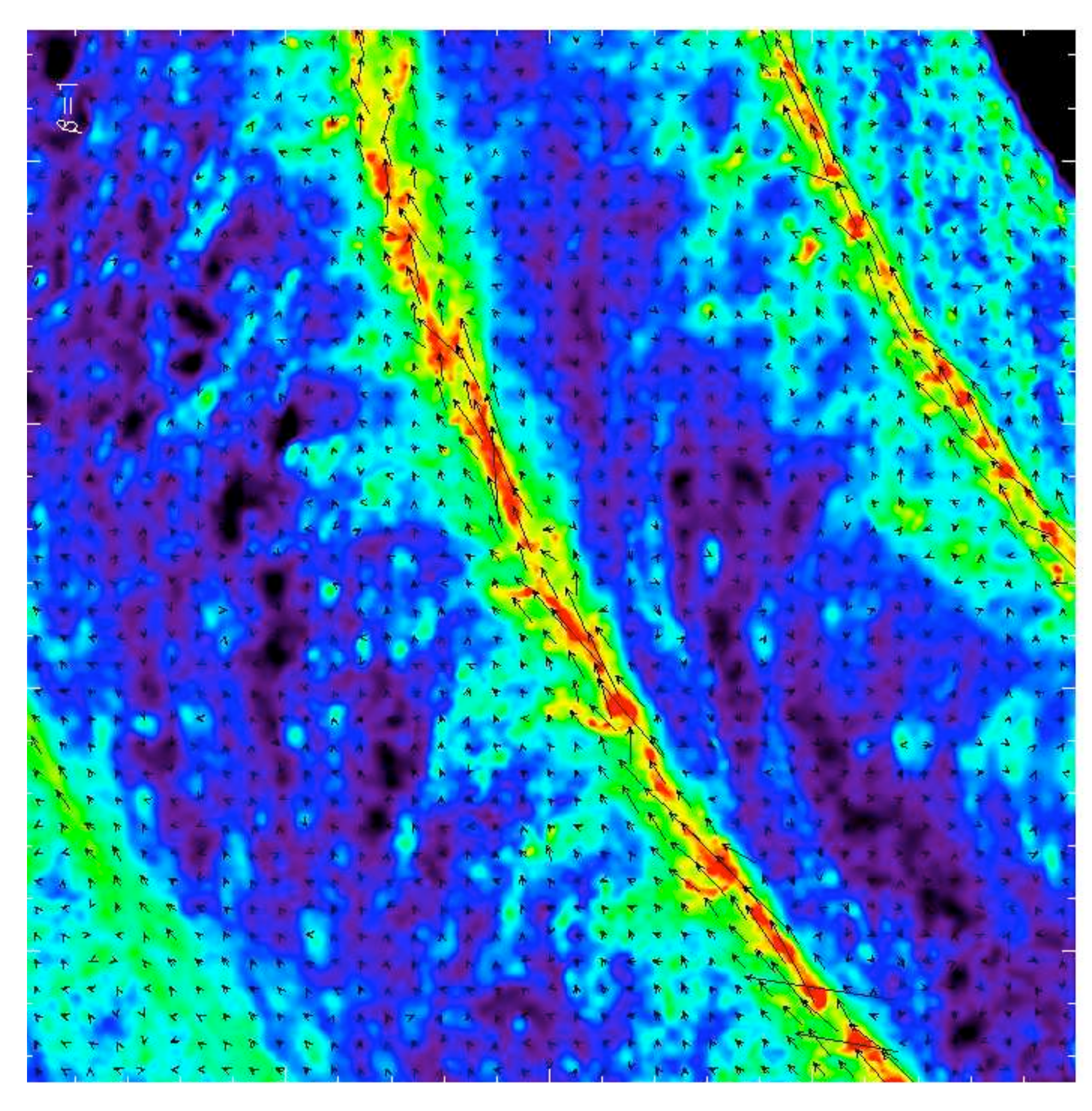}
\includegraphics[scale=0.24,angle=270]{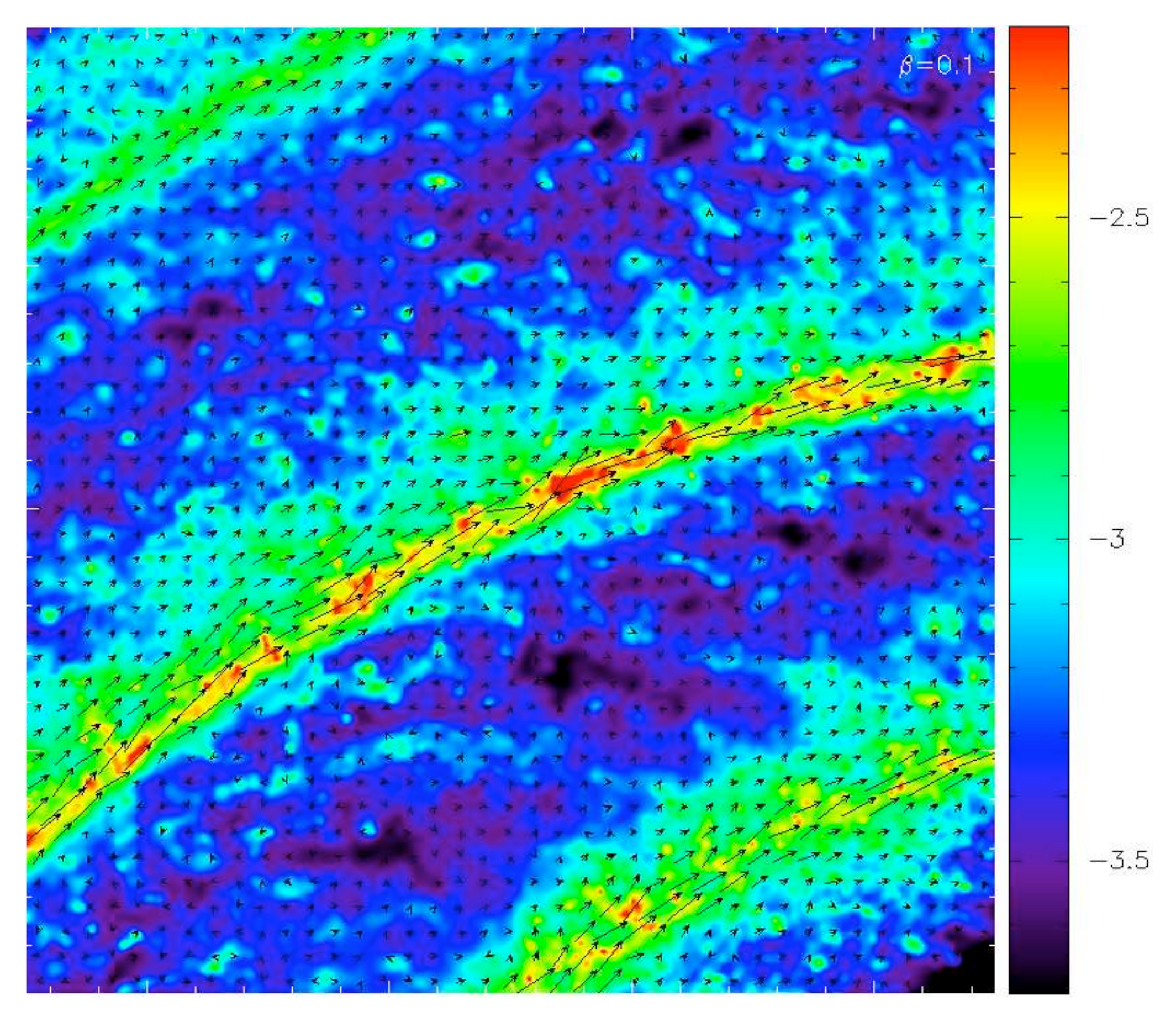}}
\caption{The spiral arm structure of a 4 kpc x 4 kpc subsection of the disk is shown when the gas is cold (100 K) for varying magnetic field strengths. The ratio of the thermal to magnetic pressure ($\beta$) is 10 (left), 1 (middle) and 0.1 (right). The right hand scale indicates the logarithmic column density (in g cm$^{-2}$) and the the vectors indicate the magnetic field integrated through $z$. There is less structure along the spiral arms for increasing field strength, and the magnetic field is more ordered.}
\end{figure}
\section{Conclusions}
For non self-gravitating galactic disks, the thermal and magnetic pressure determine the density and degree of structure along the spiral arms, and thus whether molecular clouds can form. If the ISM is assumed to be a two phase medium, the structure in the cold gas is enhanced from the pressure of the warm component, and consequently more molecular clouds are present in the inter-arm regions. The spiral shock not only affects the spiral structure, but also induces a velocity dispersion in the gas and a random component of the magnetic field.
\\
\acknowledgments Some of this work was conducted as part of the award `The formation of stars and planets: Radiation hydrodynamical and magnetohydrodynamical simulations' made under the European Heads of Research Councils and European Science Foundation EURYI (European Young Investigator) Awards scheme, supported by funds from the Participating Organisations of EURYI and the EC Sixth Framework Programme. The computations reported here were performed using the UK Astrophysical Fluids Facility (UKAFF). DJP is supported by a PPARC postdoctoral research fellowship.

\end{document}